\documentclass[journal=nalefd,manuscript=letter]{achemso}

\usepackage{amsmath}
\usepackage{amssymb}
\usepackage{textcomp}

\usepackage{graphicx}
\graphicspath{{figures/}}

\author{J.~Houel}
\affiliation{Institut Lumi{\`e}re-Mati{\`e}re, CNRS UMR5306, Universit{\'e}
Lyon 1, Universit{\'e} de Lyon, 69622 Villeurbanne CEDEX, France}
\email{julien.houel@univ-lyon1.fr}

\author{Q.\,T.~Doan}
\affiliation{Institut de Physique Nucl{\'e}aire de Lyon, CNRS UMR5822,
Universit{\'e} Lyon 1, Universit{\'e} de Lyon, 4 rue Enrico Fermi, 69622
Villeurbanne CEDEX, France}

\author{T.~Cajgfinger}
\affiliation{Institut de Physique Nucl{\'e}aire de Lyon, CNRS UMR5822,
Universit{\'e} Lyon 1, Universit{\'e} de Lyon, 4 rue Enrico Fermi, 69622
Villeurbanne CEDEX, France}

\author{G.~Ledoux}
\affiliation{Institut Lumi{\`e}re-Mati{\`e}re, CNRS UMR5306, Universit{\'e}
Lyon 1, Universit{\'e} de Lyon, 69622 Villeurbanne CEDEX, France}

\author{D.~Amans}
\affiliation{Institut Lumi{\`e}re-Mati{\`e}re, CNRS UMR5306, Universit{\'e}
Lyon 1, Universit{\'e} de Lyon, 69622 Villeurbanne CEDEX, France}

\author{A.~Aubret}
\affiliation{Institut Lumi{\`e}re-Mati{\`e}re, CNRS UMR5306, Universit{\'e}
Lyon 1, Universit{\'e} de Lyon, 69622 Villeurbanne CEDEX, France}

\author{A.~Dominjon}
\affiliation{Institut de Physique Nucl{\'e}aire de Lyon, CNRS UMR5822,
Universit{\'e} Lyon 1, Universit{\'e} de Lyon, 4 rue Enrico Fermi, 69622
Villeurbanne CEDEX, France}

\author{S.~Ferriol}
\affiliation{Institut de Physique Nucl{\'e}aire de Lyon, CNRS UMR5822,
Universit{\'e} Lyon 1, Universit{\'e} de Lyon, 4 rue Enrico Fermi, 69622
Villeurbanne CEDEX, France}

\author{R.~Barbier}
\affiliation{Institut de Physique Nucl{\'e}aire de Lyon, CNRS UMR5822,
Universit{\'e} Lyon 1, Universit{\'e} de Lyon, 4 rue Enrico Fermi, 69622
Villeurbanne CEDEX, France}

\author{M.~Nasilowski}
\affiliation{Laboratoire de Physique et d'{\'E}tude des Mat{\'e}riaux,
CNRS UMR8213, {\'E}cole Sup{\'e}rieure de Physique et de Chimie
Industrielles de la Ville de Paris, 10 Rue Vauquelin, 75231 Paris CEDEX
05, France}

\author{E.~Lhuillier}
\affiliation{Laboratoire de Physique et d'{\'E}tude des Mat{\'e}riaux,
CNRS UMR8213, {\'E}cole Sup{\'e}rieure de Physique et de Chimie
Industrielles de la Ville de Paris, 10 Rue Vauquelin, 75231 Paris CEDEX
05, France}

\author{B.~Dubertret}
\affiliation{Laboratoire de Physique et d'{\'E}tude des Mat{\'e}riaux,
CNRS UMR8213, {\'E}cole Sup{\'e}rieure de Physique et de Chimie
Industrielles de la Ville de Paris, 10 Rue Vauquelin, 75231 Paris CEDEX
05, France}

\author{C.~Dujardin}
\affiliation{Institut Lumi{\`e}re-Mati{\`e}re, CNRS UMR5306, Universit{\'e}
Lyon 1, Universit{\'e} de Lyon, 69622 Villeurbanne CEDEX, France}

\author{F.~Kulzer}
\affiliation{Institut Lumi{\`e}re-Mati{\`e}re, CNRS UMR5306, Universit{\'e}
Lyon 1, Universit{\'e} de Lyon, 69622 Villeurbanne CEDEX, France}
\email{florian.kulzer@univ-lyon1.fr}

\title{Autocorrelation analysis for the unbiased determination of
power-law exponents in single-quantum-dot blinking}

\keywords{Colloidal Quantum Dots, Photoluminescence, Power-Law Blinking,
Intensity Autocorrelation, Non-Ergodicity}

\begin{document}




\clearpage

\begin{abstract}
We present an unbiased and robust analysis method for power-law blinking
statistics in the photoluminescence of single nano-emitters, allowing us
to extract both the bright- and dark-state power-law exponents from the
emitters' intensity autocorrelation functions. As opposed to the
widely-used threshold method, our technique therefore does not require
discriminating the emission levels of bright and dark states in the
experimental intensity timetraces. We rely on the simultaneous recording
of 450 emission timetraces of single CdSe/CdS core/shell quantum dots at
a frame rate of 250\,Hz with single photon sensitivity. Under these
conditions, our approach can determine ON and OFF power-law exponents
with a precision of 3\,\% from a comparison to numerical simulations,
even for shot-noise-dominated emission signals with an average intensity
below 1 photon per frame and per quantum dot. These capabilities pave
the way for the unbiased, threshold-free determination of blinking
power-law exponents at the micro-second timescale.
\end{abstract}


Blinking, that is to say intermittent fluorescence \cite{nirmal96oct31,
banin99jan8, kuno00feb15, shimizu01may15}, is a ubiquitous feature of
the emission of nanoparticles \cite{frantsuzov08natph4519} and can have
dramatic consequences for many potential applications. For colloidal
quantum dots (QDs), blinking affects the performance of lasers
\cite{klimov00oct13}, light emitting diodes \cite{anikeeva07nanol72196}
and single photon sources \cite{lounis00oct27, michler00aug31}, to name
but a few examples. Photoluminescence (PL) intermittence manifests
itself as intensity fluctuations in the fluorescence timetrace of
nano-emitters, where highly-emitting states (ON states) are repeatedly
interrupted by poorly-emitting states (OFF states). The durations of
these alternating ON and OFF periods are found to be distributed
according to power laws for many kinds of quantum emitters
\cite{frantsuzov08natph4519}, including CdSe/CdS QDs. Under these
distributions, the probability $P_\text{\tiny ON}(t) \, dt$ of observing
an ON state duration between $t$ and $t + dt$ is governed by the
probability density
\begin{equation}\label{eq:powerlaw}
P_\text{\tiny ON}(t) = (m_\text{\tiny ON} - 1) \cdot
\theta^{m_\text{ON}-1} \cdot t^{-m_\text{ON}}
\qquad,
\end{equation}
where $m_\text{\tiny ON}$ is the power-law exponent associated with the
ON state and $\theta$ is the cut-on time of the blinking process. The
expression for the OFF-state probability density $P_\text{\tiny OFF}(t)$
can be obtained from Eq.~(1) by replacing $m_\text{\tiny
ON}$ with $m_\text{\tiny OFF}$, the corresponding exponent for the OFF
state. For colloidal QDs, power law exponents $\lesssim 2$ have been
found, which implies non-ergodicity of the ON- and OFF-state dynamics
\cite{brokmann03mar28, lutz04nov5}.

Theoretical efforts to explain power-law-like emission characteristics
started with Randall and Wilkins, who showed that the existence of
electron traps with exponentially-distributed depths explains power-law
decay of phosphorescence \cite{randall45procr184390}. As far as QDs are
concerned, their OFF states are linked to charge separation and electron
trapping \cite{efros97feb10}, meaning that similar considerations can be
applied. Power-law distributed ON times, on the other hand, are less
straightforward to account for. More elaborate models have therefore
been developed, based on spectral diffusion \cite{shimizu01may15,
tang05sep2}, fluctuating barriers \cite{kuno01jul8, kuno03mar15}, the
existence of charged ON states \cite{verberk02dec15}, spatial diffusion
\cite{margolin06advch133327}, and variations of non-radiative rates
\cite{frantsuzov05physr72155321}. However, while each of these models
reproduces a large part of the available experimental evidence, there is
still no unified approach that explains all observed properties of QD
fluorescence intermittency; as a further complication, the existing
models predict different power-law exponents. Recent experimental
results have furthermore hinted at the possibility of subtle variations
of the exponents when changing parameters like the excitation wavelength
\cite{knappenberger07nanol73869} or the excitation power
\cite{goushi09nov26, malko11dec115213}. As a consequence, an accurate
and reliable method to determine power-law exponents from experimental
data appears to be crucial for all further efforts toward a unified
understanding of the underlying physical phenomena.

Several sophisticated methods exist for the analysis of
single-nano-emitter blinking \cite{lippitz05may}. Studies of power-law
blinking usually proceed by first identifying the ON and OFF periods in
single-particle fluorescence timetraces and then adjusting
Eq.~(1) to the probability densities of the observed ON
and OFF times \cite{kuno00feb15, shimizu01may15, kuno01jul8,
sher08mar10}. The standard procedure of least-squares fitting is known
to have problems with long-tailed distributions
\cite{goldstein04eurph41255}. Thus, more suitable methods to extract
$m_\text{\tiny ON(OFF)}$ have been developed, based on
maximum-likelihood criteria and other statistical tests
\cite{goldstein04eurph41255, hoogenboom06nov28, clauset09siamr51661,
riley12may14}. Nevertheless, all these approaches still crucially depend
on a reliable distinction between ON and OFF in the emission intensity
traces, which involves establishing an acceptable intensity threshold
for a binned timetrace. The nano-emitter is thus considered to be in the
ON-state if the intensity of a time bin surpasses this threshold and to
be in the OFF-state otherwise, which is straightforward in both concept
and implementation. However, it has been shown recently
\cite{crouch10nanol101692} that the extracted $m_\text{\tiny ON}$ and
$m_\text{\tiny OFF}$ can differ by up to $30\%$, depending on the
experimental resolution (bin time) and the chosen threshold value.
Furthermore, this method obviously depends on the signal-to-noise ratio
(SNR) and thus breaks down when the signals are dominated by shot noise,
which blurs the distinction between ON and OFF levels and thus limits
the temporal resolution that can be achieved.

The change-point detection approach of Watkins \textit{et al.}
\cite{watkins05jan13} is an alternative to the threshold method: Here,
the arrival time of every photon is recorded with high temporal
resolution ($\sim 100$\,ns); a subsequent maximum-likelihood analysis
can then identify the most probable times at which
ON$\leftrightarrow$OFF transitions occurred. The bin-time bias is thus
eliminated as the technique makes the best possible use of the temporal
resolution of the data-acquisition electronics. However, a trade-off
still exists between efficiency (detecting all state changes, avoiding
false negatives) and purity (detecting only ``real" state changes,
avoiding false positives). This constraint reintroduces a user-biased
choice for the acceptable level of false positives, with a concomitant
trade-off for false negatives, in the maximum-likelihood analysis.

Two approaches have been explored for extracting $m_\text{\tiny ON}$ and
$m_\text{\tiny OFF}$ power-law blinking exponents without trying to
differentiate ON and OFF states explicitly in the timetrace
\cite{verberk03jul22, pelton04aug2}. These methods successfully recover
the power-law exponent if only one power-law process is at work, but
become ambiguous as soon as two such distributions are involved, as is
the case for QD blinking. Pelton \textit{et al.} \cite{pelton04aug2} have
analyzed the power spectrum of an ensemble of QDs to show that the
Fourier transform of their emission timetrace behaves as $1/f^{\beta}$,
where $\beta$ contains the information on both ON and OFF time periods;
so far it has not been possible to disentangle the individual
contributions of $m_\text{\tiny ON}$ and $m_\text{\tiny OFF}$.

Verberk \textit{et al.} \cite{verberk03jul22} present an analysis based
on the fluorescence intensity autocorrelation function, which makes use
of the full information contained in the delays between all pairs of
detected photons. As such, it is less sensitive to noise, can be applied
to the data at full temporal resolution, and does not require any ON/OFF
intensity threshold to be defined. However, the autocorrelation function
contains an intermixed information on $m_\text{\tiny ON}$ and
$m_\text{\tiny OFF}$; so far no general analytical expression to extract
$m_\text{\tiny ON}$ and $m_\text{\tiny OFF}$ from the autocorrelation
function has been put forward.


In this letter, we present the unbiased determination of $m_\text{\tiny
ON}$ and $m_\text{\tiny OFF}$ power-law blinking exponents of CdSe/CdS
QDs using the autocorrelation function. Our approach is robust with
respect to experimental noise and temporal resolution, allowing the
extraction of power-law exponents from fast ($2$\,ms integration time),
low-signal ($<1$ photon per frame for each QD on average) blinking data.
The method, which we here apply to the PL of single CdSe/CdS quantum
dots, does not require setting an intensity threshold for distinguishing
ON and OFF states in the experimental emission timetrace, thus removing
the potential bias \cite{crouch10nanol101692} inherent in making such a
choice. Furthermore, our technique can easily be extended to
photophysical schemes that involve more than two states and we therefore
expect it to be applicable to many different types of nano-emitters.


The fact that power-law blinking lacks a typical timescale has dramatic
consequences: To obtain complete information on the fluorescence
dynamics of single nano-emitters, the total experimental time needs to
be infinite. As a consequence, experimental autocorrelation functions,
even of one and the same nano-emitter, recorded at different times can
deviate from each other significantly. This is not necessarily due to
any change in the blinking behavior (the underlying power-law exponents
themselves), but rather an intrinsic signature of the non-ergodicity
(statistical aging) of luminescence that is governed by power-laws
\cite{brokmann03mar28, lutz04nov5}. We therefore record a large number
of single QD fluorescence timetraces simultaneously so that we can
perform a statistical analysis of the corresponding autocorrelation
functions; a subsequent comparison to numerical simulations identifies
the best-fit power-law exponents with high specificity.


\begin{figure}[tbp]
\includegraphics[width=8.46cm]{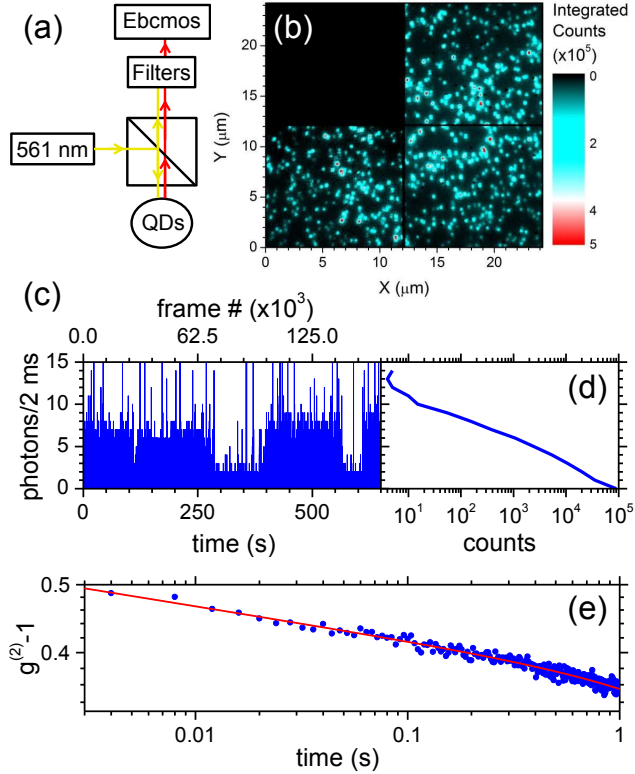}
\caption{(a) Scheme of the experimental setup. Single QDs are excited at
561\,nm with a continuous-wave laser. Photoluminescence is collected
through the excitation objective and directed onto the ebCMOS camera.
Scattered laser light is suppressed by long-pass filters. (b)
Position-dependent integrated photon counts per pixel on a false-color
scale; the total acquisition time was 660\,s. (The black square in the
top left is due to one of the 4 camera quadrants having been turned off
during the measurement.) (c) An example of a single QD timetrace
extracted from (b). (d) Distribution of the counts of the QD timetrace:
no global threshold can be established to discriminate between the ON
and OFF states at full temporal resolution (2\,ms). (e) Autocorrelation
function of the data in (c) (blue dots) and the corresponding fit of
Eq.~(3) (red line) with parameters $A=0.37$, $B=0.047$
and $C=0.049$.}
\label{figure1}
\end{figure}


The experimental setup used to record the timetraces is a home-built
wide-field microscope coupled to a high-frame-rate ebCMOS camera
\cite{barbier11aug21, doan12procs8436, guerin12dec11} with high fidelity
single photon counting capabilities, see Fig.~\ref{figure1}\,a. The
CdSe/CdS core (3\,nm)/shell(8\,nm) QDs have an emission maximum centered
at 597\,nm and spin-coated onto a glass slide from a 90/10 hexane/octane
solution. QD luminescence is excited by a 561\,nm solid state laser with
an intensity of 200\,W/cm$^{2}$ in the center of the laser spot. The
emission of individual QDs is collected by a $60 \times$, $NA=1.35$
oil-immersion objective and is redirected onto the ebCMOS camera with a
plano-convex lens of 1\,m focal length, resulting in $333\times$
magnification. The overall detection efficiency of the apparatus is
around 3\%. (Further details on the setup and the QD samples are
available in the Supporting Information, Sections 1 and 2.)

We have recorded the fluorescence of 450 single QDs simultaneously at a
frame rate of $250$\,Hz with a total integration time of 660 seconds. It
is worth mentioning that this frame rate is achieved on the full ebCMOS
camera chip of $800 \times 800$ pixels. To our knowledge, this is the
first report of such a large number of single QD timetraces recorded
simultaneously at such a high frame rate and with the single photon
sensitivity. To validate our method beyond standard conditions (slow
acquisition and relatively high SNR), we deliberately kept the
excitation power to a minimum, resulting in single-QD timetraces with
average count rates of $\sim1$ photon per frame. Such low-level signals
can be recorded with the ebCMOS sensor thanks to its ultra-small dark
noise of less than $0.02$ photons/QD/frame on average (see Supporting
Information, Fig.~S9). Fig.~\ref{figure1}\,b shows the integrated image
of the emission of 450 individual QDs, to which a pattern recognition
algorithm was applied to locate the positions of the QDs (see Supporting
Information, Section 3). The signal of each QD is then extracted from
the sequence of images as a $165\,000$-frame timetrace, an example of
which is shown in Fig.~\ref{figure1}\,c. As can be seen in
Fig.~\ref{figure1}\,d, the distribution of photon counts as commonly
used in threshold-based methods \cite{goldstein04eurph41255,
hoogenboom06nov28, clauset09siamr51661, riley12may14} does not allow for
the discrimination between ON and OFF states.

To analyze the single-QD timetraces, their fluorescence intensity
autocorrelation functions $g^{(2)}(\tau)$ are calculated according to:
\begin{equation}\label{eq:g2}
g^{(2)}(\tau)=\frac{\bigl\langle I(t) \, I(t+\tau) \bigr\rangle}
{\bigl\langle I(t) \bigr\rangle^{2}}
\qquad,
\end{equation}
where $I(t)$ is the intensity (counts per timebin) at time $t$ and
$\bigl \langle \cdot \bigr \rangle$ represents time averages;
Fig.~\ref{figure1}\,e shows an example of a single-QD autocorrelation
function. Power-law blinking with exponents $m<2$ lead to timetraces
that are dominated by long events whose duration is of the same order of
magnitude as the total measurement time \cite{bardou02levylasercooling}.
As a consequence, the normalization factor $\langle I(t) \rangle^2$ in
Eq.~(2) does not tend toward a well-defined long-time limit.
The experimental autocorrelation functions therefore show significant
variation from one QD to the next, and even if one and the same QD
is probed several times under identical experimental conditions.
Nonetheless, the autocorrelation functions exhibit a well-defined
general shape for almost all (more than 95\%) of the 450 QDs we studied:
a power-law decay modulated by an exponential cut-off, in accordance
with earlier reports \cite{verberk03jul22}. The red line in
Fig.~\ref{figure1}\,e shows a fit of the autocorrelation with the
following equation:
\begin{equation}\label{eq:fitfunc}
f(t) = A \, t^{-C} \, \exp \bigl( -B t \bigr)
\quad,
\end{equation}
where $A$ represents the autocorrelation contrast, $B$ the cut-off time
and $C$ is the power-law exponent of the autocorrelation function; $C$
is equal to $2-m$ if only one of the two states has lifetimes governed
by a power law with exponent $m$\cite{verberk03jul22,
bardou02levylasercooling}. Generally speaking, the decay of an
autocorrelation function represents a loss of information about the
state of the emitter: As time progresses, it becomes increasingly likely
that transitions occur, and at long times one can only make general
statistical predictions that are independent of the emitter's state at
time $t=0$. We can therefore surmise that the fit parameter $C$ will be
linked to the combined contributions of the $m_\text{\tiny ON}$ and
$m_\text{\tiny OFF}$ distributions, given that both types of transitions
are stochastic in nature and hence lead to information loss. The
autocorrelation contrast $A$ is influenced by the relative duration of
the ON/OFF periods \cite{lippitz05may}; traces dominated by long OFF
periods have higher correlation contrasts than those of an emitter that
is mostly in the ON state. The exponential cut-off rate given by
parameter $B$, a phenomenological addition to the fit function
\cite{verberk03jul22}, may be attributable, at least partially, to the
finite measurement time.


\begin{figure}[tbp]
\includegraphics[width=8.46cm]{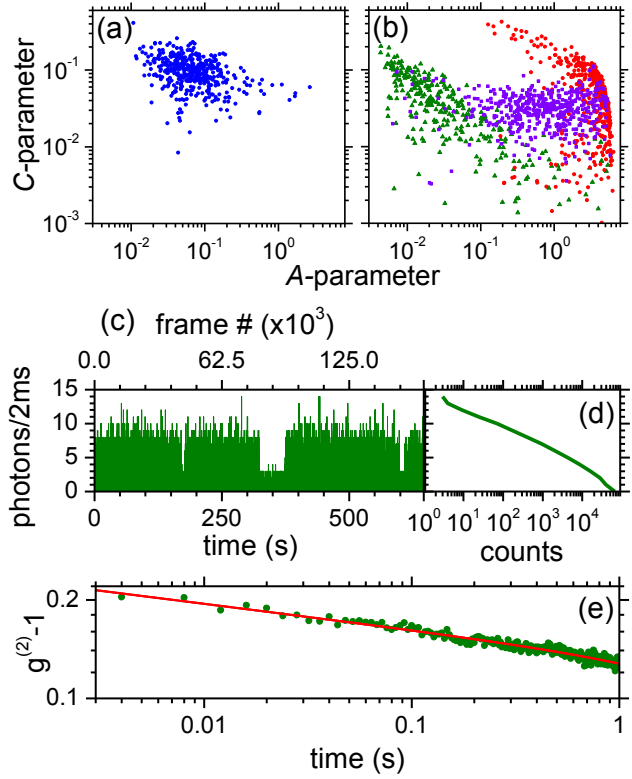}
\caption{(a) 2D distribution of the $(A,C)$ couples resulting from
fitting Eq.~(3) to the experimental autocorrelation
functions. Our analysis relies on reproducing this 2D distribution with
simulated power-law blinking timetraces that are subjected to the same
autocorrelation analysis. (b) Three different $(A,C)$ distributions
obtained after fitting the autocorrelation functions of simulated traces
for three different sets of ($m_\text{\tiny ON}, m_\text{\tiny OFF}$)
exponents. Green triangles corresponds to $(1.5,1.7)$, violet squares to
$(1.7, 1.7)$ and red dots to $(1.7, 1.5)$. Every pair of exponents
generates its own 2D distribution in the $(A, C)$ space. (c) Example of
a simulated timetrace with $(m_\text{\tiny ON}=1.80, m_\text{\tiny
OFF}=1.95)$ power-law exponents. (d) Distribution of the photon counts
of the timetrace in (c). As for the experimental data, no global
threshold can be established for discriminating ON and OFF states. (e)
The corresponding autocorrelation, fitted (red line) by
Eq.~(3) with adjusted parameters $A=0.13$, $B=0.056$ and
$C=0.079$.}
\label{figure2}
\end{figure}


Based on the above heuristic arguments, we conclude that the combination
of parameters $C$ and $A$ may contain sufficient information to unravel
the contributions of $m_\text{\tiny ON}$ and $m_\text{\tiny OFF}$, even
in the absence of a general analytical formula relating the fit
parameters to the power-law exponents. (The cut-off parameter $B$ serves
as a consistency check, see Supporting Information, Section 8.) Due to
the non-ergodicity of power-law blinking, we expect to find a broad
distribution of the two parameters in $(A, C)$ space;
Fig.~\ref{figure2}\,a shows that this is indeed the case for the 450
experimental autocorrelation functions. The assumption at the heart of
our subsequent analysis is that this 2D distribution of the $(A, C)$
parameters corresponds to one and only one ($m_\text{\tiny ON},
m_\text{\tiny OFF}$) pair of blinking exponents. To validate this
assumption, we have simulated 450 single QD timetraces with ON and OFF
periods distributed according to power laws with exponents
$(m_\text{\tiny ON}, m_\text{\tiny OFF})$ (further details of the
simulations and the fitting procedure are given in the Supporting
Information, Sections 4 to 7). Fig.~\ref{figure2}\,c shows an example of
a simulated timetrace for $(m_\text{\tiny ON}, m_\text{\tiny OFF})=(1.8,
1.95)$ and Fig.~\ref{figure2}\,e presents the corresponding
autocorrelation. For every $(m_\text{\tiny ON}, m_\text{\tiny OFF})$
couple, the 450 simulated autocorrelation functions are fitted with
Eq.~(3), yielding the 2D distribution of $A$ and $C$ in
each case.


\begin{figure}[tbp]
\includegraphics[clip,width=12cm]{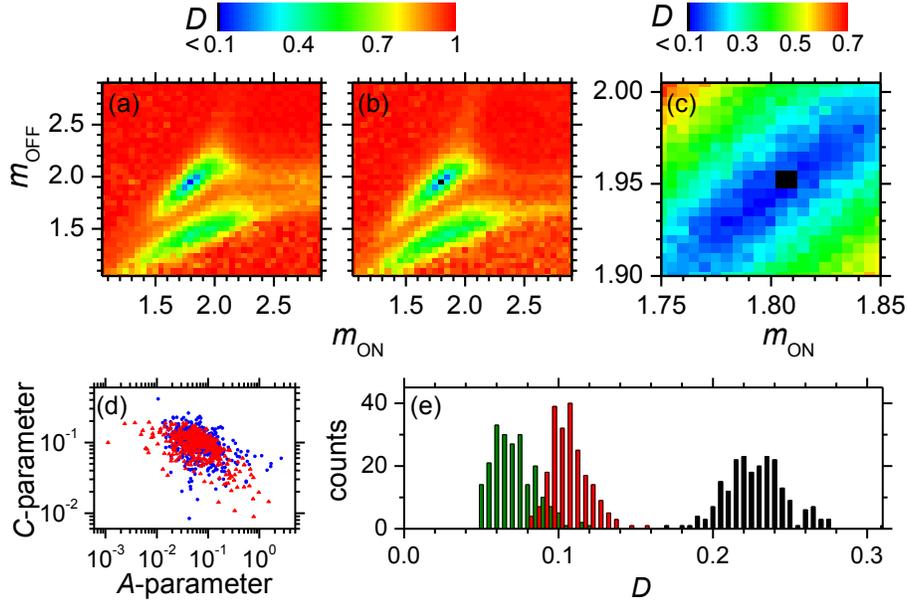}
\caption{(a) Low-resolution comparison of simulations to experimental
data with a 2D Kolmogorov-Smirnov (K-S) test. The K-S parameter $D$
(color scale) is represented as a function of the $m_\text{\tiny ON}$
and $m_\text{\tiny OFF}$ exponents used in the simulations. There is a
single $(m_\text{\tiny ON}, m_\text{\tiny OFF})$ couple, $(1.80, 1.95)$,
that minimizes the $D$-parameter, corresponding to the best agreement
between experimental and simulated $(A,C)$ distributions. (b) Same
comparison as in (a), now with a simulated set for $(m_\text{\tiny
ON}=1.80, m_\text{\tiny OFF}=1.95)$ replacing the experimental data; all
features of the original contour plot (a) are reproduced. (c)
High-resolution exploration of the area of minimal $D$ from (a),
yielding more accurate optimum values of $(m_\text{\tiny ON}=1.805,
m_\text{\tiny OFF}=1.955)$ $\pm3$\%. (d) 2D distributions of $A$ and $C$
for the data (blue dots, same as in Fig.~\ref{figure2}\,a) and the
best-fit simulation $(m_\text{\tiny ON}=1.805, m_\text{\tiny
OFF}=1.955)$ (red triangles). (e) Reproducibility and distinctiveness of
$D$: The red histogram shows the distribution found for $D$ when
comparing the experimental $(A,C)$ distribution to 215 different
analysis runs for the previously-determined optimum couple
$(m_\text{\tiny ON}=1.805, m_\text{\tiny OFF}=1.955)$, while the black
bars represent the analogous distribution for $(m_\text{\tiny ON}=1.85,
m_\text{\tiny OFF}=2.00)$, the second-lowest pixel in the contour plot
in (a). The green histogram corresponds to a null-hypothesis
calibration, for which one simulation run for $(m_\text{\tiny ON}=1.805,
m_\text{\tiny OFF}=1.955)$ is compared to 215 additional runs for the
very same pair of parameters.}
\label{figure3}
\end{figure}


Three examples of such simulated distributions are plotted in
Fig.~\ref{figure2}\,b for $(m_\text{\tiny ON}, m_\text{\tiny
OFF})=(1.5,1.7)$, $(1.7,1.7)$ and $(1.7,1.5)$. As expected, the
distributions for each $(m_\text{\tiny ON}, m_\text{\tiny OFF})$ pair
are spread over a large area in $(A,C)$ space, meaning that correctly
identifying the underlying power-law exponents requires studying a
statistically significant number of single QDs (see Supporting
Information, Section 13). Given a large-enough data set, we can test
whether a single $(m_\text{\tiny ON},m_\text{\tiny OFF})$ couple can be
identified as the ''best fit`` for describing the experimental data of
Fig.~\ref{figure2}\,a. To this end, we use a 2D Kolmogorov-Smirnov (K-S)
statistical test \cite{peacock83monno202615, fasano87mar1}, which
compares the 2D $(A,C)$ distributions of two different data sets,
yielding a parameter $D$ that quantifies the mismatch between the two
distributions: $D \in [0,1]$, where $D=0$ would correspond to prefect
overlap. In total, we have tested 1444 different $(m_\text{\tiny ON},
m_\text{\tiny OFF})$ combinations ranging from $(1.05, 1.05)$ to $(2.9,
2.9)$, covering more than the spread of values reported in the
literature \cite{hoogenboom06nov28, knappenberger07nanol73869,
frantsuzov08natph4519, mahler08natma7659, orrit10photo, riley12may14}.
That is to say, we have simulated 450 single-QD timetraces for each
$(m_\text{\tiny ON}, m_\text{\tiny OFF})$ couple, determined the
corresponding 2D distribution in $(A,C)$ space and calculated the K-S
parameter $D$ with respect to the experimental data of
Fig.~\ref{figure2}\,a. The 2D contour plot in Fig.~\ref{figure3}\,a
shows the resulting values of $D$ on a color scale as a function of
$m_\text{\tiny ON}$ and $m_\text{\tiny OFF}$; the high contrast of $D$
spans variations of one order of magnitude, from $D\simeq0.1$ to $1$.
There is an isolated, well-defined minimum of $D \lesssim 0.1$ at $(1.8,
1.95)$, indicating that a singular, narrowly-delimited combination of
exponents optimizes the overlap between the experimental data and
simulations based on the power-law model of Eq.~(1). A
high-resolution contour plot of the parameter space around the minimum
of $D$ can be seen in Fig.~\ref{figure3}\,c. For this particular
ensemble of CdSe/CdS QDs, we thus find best-fit blinking exponents of
$(m_\text{\tiny ON}=1.805, m_\text{\tiny OFF}=1.955)$ for the pixel with
minimum $D$; the corresponding simulated $(A,C)$ distribution is
compared to the experimental data in Fig.~\ref{figure3}\,d.

After having shown that our approach can identify the optimal
$(m_\text{\tiny ON}, m_\text{\tiny OFF})$ couple with high specificity,
we now discuss to what extent the autocorrelation analysis allows us to
judge whether the underlying hypothesis itself -- QD blinking is
governed by power-law distributed probabilities, Eq.~(1)
-- is justified. To explore this issue, we took a simulated data set for
$(m_\text{\tiny ON}=1.805, m_\text{\tiny OFF}=1.955)$, i.\,e., an
ensemble of timetraces for which we know the null hypothesis to be true,
and we subjected this set to the same analysis as the experimental data.
We can thus identify the behavior of $D$ that corresponds to genuine
power-law blinking and quantify the degree of variation in $D$ that is
inherent in repeatedly probing the same power-law distributions with
limited sample sizes and measurement times. As can be seen in
Fig.~\ref{figure3}\,b, the resulting ''ideal`` contour plot agrees very
well with the experimental one of Fig.~\ref{figure3}\,a, down to the
shape of the faint offshoots observed for the main and secondary minima.
However, the values of $D$ are slightly lower in the minimum regions of
Fig.~\ref{figure3}\,b, although this is barely noticeable given the
color scale. We further investigated this feature by subjecting both the
real and the idealized (simulated) data to 215 different analysis runs
for the previously identified optimum parameters $(m_\text{\tiny
ON}=1.805, m_\text{\tiny OFF}=1.955)$. Each analysis run is based on a
new seed of the random number generator and therefore produces its own
simulated $(A,C)$ distribution, to which both data sets (real and
idealized) are then compared with the K-S test. The
simulation-simulation analyses thus yield the distribution of $D$ values
that can be expected for idealized power-law blinking, which, as is
shown in Fig.~\ref{figure3}\,e (green histogram), has its mean value at
$D_\text{sim} = 0.074$ with a standard deviation of $\sigma_\text{sim} =
0.014$. The experiment-simulation analysis runs, on the other hand,
produce a roughly Gaussian-shaped histogram (red) with mean value
$D_\text{exp} = 0.107$ and standard deviation $\sigma_\text{exp} =
0.013$. There is about 20\% overlap between the experiment-simulation
and the simulation-simulation distributions, with the $D$ values for the
experimental data being larger in general. This means that the data, on
average, tends to agree slightly less well with simulations than can be
expected from the variations between equivalent simulation-simulation
analysis runs. Nevertheless, the large overlap means that there is no
reason to reject the null hypothesis at the base of our analysis, which
supposed that the blinking behavior of all the investigated QDs can be
modeled by a power law with a single $(m_\text{\tiny ON}, m_\text{\tiny
OFF})$ combination. The remaining small offset between $D_\text{exp}$
and $D_\text{sim}$ might be due to an aspect of the particles'
photophysics that is not incorporated in our model. For example, small
inhomogeneities may be present in the investigated sample of 450 QDs as
far as power-law exponents, the exciton emission rates and/or the ratios
between bright and dark state emission efficiencies are concerned.

The black histogram in Fig.~\ref{figure3}\,e is the result of the
experiment-simulation comparison for $(m_\text{\tiny ON}=1.85,
m_\text{\tiny OFF}=2.00)$, which corresponds to the pixel with the
second-lowest $D$ in the contour plot of Fig.~\ref{figure3}\,a. There is
strictly no overlap with the $D$ distribution for the optimum fit
parameters (red histogram), illustrating once more the specificity of
the autocorrelation analysis. In fact, as is detailed in the Supporting
Information (Sections 9 and 11), we find that all 8 nearest-neighbor
pixels in Fig.~\ref{figure3}\,a exhibit distributions whose maxima
differ by at least $6\sigma$ from the mean value of $D = 0.107$ of the
optimum-solution histogram (red); where $\sigma$ stands for the largest
standard deviation of the compared histograms (worst case scenario). We
therefore conclude that we are able to extract the power-law exponents
with an absolute precision of $\pm0.05$ ($\pm3$\%) at $6\sigma$
specificity. The combination of $m_\text{\tiny ON}=1.805$ with an almost
10\% larger $m_\text{\tiny OFF}=1.955$ indicates that these QDs spend
most of the time in the ON state under continuous illumination, a
typical feature of such large-shell CdSe/CdS QDs
\cite{mahler08natma7659, canneson14jan9}. It is particularly noteworthy
that $m_\text{\tiny OFF}$ approaches the critical threshold of $2$,
above which the average duration of the OFF periods becomes finite. The
power-law exponent of the ON periods, on the other hand, is associated
with an infinite average length; overall, this leads to a favorable
interplay of ON versus OFF periods in the photoluminescence of this type
of QD.

To complete the discussion of our technique, we now address its
robustness with respect to two critical factors. First, we consider the
influence of the ON/OFF intensity contrast. OFF states can still be
moderately emissive (``dim'' instead of completely dark), which makes it
harder to distinguish them from the ON states. In fact, residual OFF
state emission manifests itself in the contour plot of
Fig.~\ref{figure3}\,a, which shows, besides the global minimum of
$D=0.1$, as a second domain (green) of relatively low $D$ values around
$0.4$. This secondary minimum arises due to the relatively high quantum
yield of the dark state for this type of QD, reaching $10$\% of the
bright state emission. We show in the Supporting Information (Section
15) that this region shifts as a function of the dark state emissivity
and tends to vanish if this emissivity drops below $\sim 0.1\%$ of the
efficiency of the bright state. With regard to more emissive ``dark''
states, we verified (see Supporting Information, Section 15) that our
technique maintains a precision of $\pm0.05$ (under the experimental
conditions discussed in this work) as long as dark state efficiencies
stay below $50$\% of the bright states. As a consequence, the approach
is also suitable for analyzing recently developed types of giant-shell
\cite{galland13nanol13321, javaux13natna8206, canneson14jan9} or alloyed
QDs \cite{wang09jun4}, both of which having a high dark-state-emission
efficiency.

The second important benchmark is the interplay between count rate,
temporal resolution and residual uncertainty for the power-law
exponents, which is linked to the sensitivity of the $D$ parameter. As
discussed above, we are able to extract the power-law exponents with an
absolute precision of $\pm0.05$ ($\pm3$\%) at $6\sigma$ specificity. It
is worth noting that this precision is achieved with
shot-noise-dominated timetraces, well below saturation of the QD
emission. Such minimally-invasive conditions are preferable to
approaches that require high count rates to discriminate between ON and
OFF states, and hence high excitation intensities that may influence the
blinking parameters \cite{malko11dec115213, goushi09nov26} and can
furthermore lead to photobleaching. As far as the temporal resolution is
concerned, our method can extract blinking power-law exponents for
timetraces with only 0.1 photons/QD/frame on average, with a reasonable
acquisition time $T_\text{max}=66$\,s with $3\%$ precision ($\pm0.05$)
at $6\sigma$ specificity (see Supporting Information, Section 14). This
robustness of our method against noise may allow blinking studies at up
to 100\,kHz (10\,\textmu s resolution), one order of magnitude faster
than what has been demonstrated with change-point detection
\cite{watkins05jan13}. Verifying power-law behavior at the fastest
possible timescale will be useful to elucidate the role of the cut-on
time, $\theta$ in Eq.~(1). Taking a pragmatic point of
view, this cut-on time can be equated with the experimental temporal
resolution; nevertheless, a more fundamental approach can be expected to
improve our understanding of QD photophysics, for example if a timescale
can be identified at which the power-law behavior breaks down.


In conclusion, we have presented a technique to determine unbiased
power-law exponents of blinking CdSe/CdS core/shell QDs with a precision
of 3\,\% at $6 \sigma$ specificity. To our knowledge, this constitutes
the first approach for extracting the full set of blinking parameters
from experimental autocorrelation functions, bypassing the need of
introducing a possibly-biased ON/OFF threshold. Our autocorrelation
analysis is robust in the presence of noise and intrinsically free from
timebin-dependent thresholding artifacts. As such, the method is capable
of determining $m_\text{\tiny ON}$ and $m_\text{\tiny OFF}$ from
timetraces dominated by shot noise, which are untreatable by other
methods. We thus can extract the power-law exponents from ultra-low
signal data ($\sim 0.1$\,photon/frame/QD) with a precision of 3$\%$,
which offers the perspective of threshold-free blinking analysis at the
micro-second timescale.


\paragraph{Acknowledgment.}

We acknowledge technical support by J.~Margueritat, J.-F.~Sivignon,
Y.~Guillin, and the Lyon center for nano-opto technologies (NanOpTec).
This research was supported by the Programme Avenir Lyon
Saint-{\'E}tienne (ANR-11-IDEX-0007) of Universit{\'e} de Lyon, within
the program ``Investissements d'Avenir'' operated by the French National
Research Agency (ANR). J.~Houel thanks the F{\'e}d{\'e}ration de
Recherche Andr{\'e} Marie Amp{\`e}re (FRAMA) for financial support. This
work was performed in the context of the European COST Action MP1302
NanoSpectroscopy.


\paragraph{Supporting Information Available.}

Extensive supporting information on experimental techniques, numerical
simulations, and further capability benchmarks of our approach is
available free of charge via the Internet at http://pubs.acs.org.


\providecommand{\textit}[1]{#1}
\providecommand*\mcitethebibliography{\thebibliography}
\csname @ifundefined\endcsname{endmcitethebibliography}
  {\let\endmcitethebibliography\endthebibliography}{}


\end{document}